\documentclass[aps,prl,twocolumn,balance,groupedaddress,amsmath,reprint]{revtex4-1}
\usepackage[latin1]{inputenc}    
\usepackage{graphicx}   
\usepackage{xspace}
\usepackage{color}
\usepackage{amsfonts}
\providecommand{\abs}[1]{\lvert#1\rvert}

\begin{document}


\newcommand{\addSG}[1]{\textcolor{magenta}{#1}} 
\newcommand{\mpar}[1]{\marginpar{\small \it \textcolor{magenta}{#1}}}
\newcommand*{\Psw}{\ensuremath{P_\mathrm{sw}}\xspace}
\newcommand*{\tpw}{\ensuremath{t_\mathrm{pw}}\xspace}
\newcommand*{\rotq}{\ensuremath{\theta}\xspace}
\newcommand*{\Aq}{\ensuremath{A_\mathrm{Q}}\xspace}
\newcommand*{\Ai}{\ensuremath{A_\mathrm{I}}\xspace}
\newcommand*{\Hint}{\ensuremath{\hat{H}_\mathrm{int}}\xspace}
\newcommand*{\Hsub}{\ensuremath{\hat{H}_\mathrm{sub}}\xspace}
\newcommand*{\Hqb}{\ensuremath{\hat{H}_\mathrm{qb}}\xspace}
\newcommand*{\Htls}{\ensuremath{\hat{H}_\mathrm{TLS}}\xspace}
\newcommand*{\siX}{\ensuremath{\hat{\sigma}_\mathrm{x}}\xspace}
\newcommand*{\siY}{\ensuremath{\hat{\sigma}_\mathrm{y}}\xspace}
\newcommand*{\siZ}{\ensuremath{\hat{\sigma}_\mathrm{z}}\xspace}
\newcommand*{\siXq}{\ensuremath{\hat{\sigma}_\mathrm{x}^\mathrm{qb}}\xspace}
\newcommand*{\siYq}{\ensuremath{\hat{\sigma}_\mathrm{y}^\mathrm{qb}}\xspace}
\newcommand*{\siZq}{\ensuremath{\hat{\sigma}_\mathrm{z}^\mathrm{qb}}\xspace}
\newcommand*{\siXt}{\ensuremath{\hat{\sigma}_\mathrm{x}^\mathrm{TLS}}\xspace}
\newcommand*{\siYt}{\ensuremath{\hat{\sigma}_\mathrm{y}^\mathrm{TLS}}\xspace}
\newcommand*{\siZt}{\ensuremath{\hat{\sigma}_\mathrm{z}^\mathrm{TLS}}\xspace}
\newcommand*{\siXs}{\ensuremath{\hat{\sigma}_\mathrm{x}^\mathrm{sub}}\xspace}
\newcommand*{\siYs}{\ensuremath{\hat{\sigma}_\mathrm{y}^\mathrm{sub}}\xspace}
\newcommand*{\siZs}{\ensuremath{\hat{\sigma}_\mathrm{z}^\mathrm{sub}}\xspace}
\newcommand*{\Isq}{\ensuremath{I_\mathrm{b}}\xspace}
\newcommand*{\PhiQ}{\ensuremath{\Phi_\mathrm{qb}}\xspace}
\newcommand*{\fqb}{\ensuremath{\omega_\mathrm{qb}}\xspace}
\newcommand*{\ftls}{\ensuremath{f_\mathrm{TLS}}\xspace}
\newcommand*{\fosc}{\ensuremath{f_\mathrm{osc}}\xspace}
\newcommand*{\dph}{\ensuremath{\delta\Phi}\xspace}
\newcommand*{\df}{\ensuremath{\delta\!f}\xspace}
\newcommand*{\dpint}{\ensuremath{\delta\Phi_\mathrm{int}}\xspace}
\newcommand*{\TphN}{\ensuremath{T_{\varphi,\mathrm{N}}}\xspace}
\newcommand*{\Tph}[1]{\ensuremath{T_{\varphi,\mathrm{#1}}}\xspace}
\newcommand*{\tp}{\ensuremath{t_\mathrm{p}}\xspace}
\newcommand*{\fRabi}{\ensuremath{f_\mathrm{Rabi}}\xspace}
\newcommand*{\Ian}{\ensuremath{I_\mathrm{antenna}^\mathrm{mw}}\xspace}
\newcommand*{\Imr}{\ensuremath{I_\mathrm{r}^\mathrm{mw}}\xspace}
\newcommand*{\emd}{\ensuremath{\varepsilon^\mathrm{mw}_\mathrm{direct}}\xspace}
\newcommand*{\emt}{\ensuremath{\varepsilon^\mathrm{mw}}\xspace}

\newcommand{\ket}[1]{\vert #1\rangle} 
\newcommand{\bra}[1]{\langle #1\vert} 
\newcommand{\braket}[1]{\langle #1\rangle} 

\newcommand*{\PhiX}{\ensuremath{\Phi_\mathrm{X}}\xspace}
\newcommand*{\PhiZ}{\ensuremath{\Phi_\mathrm{Z}}\xspace}
\newcommand*{\fX}{\ensuremath{f_\mathrm{x}}\xspace}
\newcommand*{\fZ}{\ensuremath{f_\mathrm{z}}\xspace}
\newcommand*{\Ax}{\ensuremath{A_\mathrm{x}}\xspace}
\newcommand*{\Az}{\ensuremath{A_\mathrm{z}}\xspace}

\newcommand*{\TF}{\ensuremath{T_{\varphi F}}\xspace}
\newcommand*{\TE}{\ensuremath{T_{\varphi E}}\xspace}
\newcommand*{\GF}{\ensuremath{\Gamma_{\varphi F}}\xspace}
\newcommand*{\GE}{\ensuremath{\Gamma_{\varphi E}}\xspace}

\newcommand*{\GSs}{\ensuremath{\,\mathrm{GS/s}\xspace}}
\newcommand*{\mPh}{\ensuremath{\,\mathrm{m}\Phi_0}\xspace}
\newcommand*{\uPh}{\ensuremath{\,\mu\Phi_0}\xspace}

\newcommand*{\um}{\ensuremath{\,\mu\mathrm{m}}\xspace}
\newcommand*{\nm}{\ensuremath{\,\mathrm{nm}}\xspace}
\newcommand*{\mm}{\ensuremath{\,\mathrm{mm}}\xspace}
\newcommand*{\m}{\ensuremath{\,\mathrm{m}}\xspace}
\newcommand*{\sqm}{\ensuremath{\,\mathrm{m}^2}\xspace}
\newcommand*{\sqmm}{\ensuremath{\,\mathrm{mm}^2}\xspace}
\newcommand*{\squm}{\ensuremath{\,\mu\mathrm{m}^2}\xspace}
\newcommand*{\psqm}{\ensuremath{\,\mathrm{m}^{-2}}\xspace}
\newcommand*{\psqmV}{\ensuremath{\,\mathrm{m}^{-2}\mathrm{V}^{-1}}\xspace}
\newcommand*{\cm}{\ensuremath{\,\mathrm{cm}}\xspace}

\newcommand*{\nF}{\ensuremath{\,\mathrm{nF}}\xspace}
\newcommand*{\pF}{\ensuremath{\,\mathrm{pF}}\xspace}
\newcommand*{\pH}{\ensuremath{\,\mathrm{pH}}\xspace}

\newcommand*{\emob}{\ensuremath{\,\mathrm{m}^2/\mathrm{V}\mathrm{s}}\xspace}
\newcommand*{\edos}{\ensuremath{\,\mu\mathrm{C}/\mathrm{cm}^2}\xspace}
\newcommand*{\mbar}{\ensuremath{\,\mathrm{mbar}}\xspace}

\newcommand*{\A}{\ensuremath{\,\mathrm{A}}\xspace}
\newcommand*{\mA}{\ensuremath{\,\mathrm{mA}}\xspace}
\newcommand*{\nA}{\ensuremath{\,\mathrm{nA}}\xspace}
\newcommand*{\pA}{\ensuremath{\,\mathrm{pA}}\xspace}
\newcommand*{\fA}{\ensuremath{\,\mathrm{fA}}\xspace}
\newcommand*{\uA}{\ensuremath{\,\mu\mathrm{A}}\xspace}

\newcommand*{\Ohm}{\ensuremath{\,\Omega}\xspace}
\newcommand*{\kOhm}{\ensuremath{\,\mathrm{k}\Omega}\xspace}
\newcommand*{\MOhm}{\ensuremath{\,\mathrm{M}\Omega}\xspace}
\newcommand*{\GOhm}{\ensuremath{\,\mathrm{G}\Omega}\xspace}

\newcommand*{\Hz}{\ensuremath{\,\mathrm{Hz}}\xspace}
\newcommand*{\kHz}{\ensuremath{\,\mathrm{kHz}}\xspace}
\newcommand*{\MHz}{\ensuremath{\,\mathrm{MHz}}\xspace}
\newcommand*{\GHz}{\ensuremath{\,\mathrm{GHz}}\xspace}
\newcommand*{\THz}{\ensuremath{\,\mathrm{THz}}\xspace}

\newcommand*{\K}{\ensuremath{\,\mathrm{K}}\xspace}
\newcommand*{\mK}{\ensuremath{\,\mathrm{mK}}\xspace}

\newcommand*{\kV}{\ensuremath{\,\mathrm{kV}}\xspace}
\newcommand*{\V}{\ensuremath{\,\mathrm{V}}\xspace}
\newcommand*{\mV}{\ensuremath{\,\mathrm{mV}}\xspace}
\newcommand*{\uV}{\ensuremath{\,\mu\mathrm{V}}\xspace}
\newcommand*{\nV}{\ensuremath{\,\mathrm{nV}}\xspace}

\newcommand*{\eV}{\ensuremath{\,\mathrm{eV}}\xspace}
\newcommand*{\meV}{\ensuremath{\,\mathrm{meV}}\xspace}
\newcommand*{\ueV}{\ensuremath{\,\mu\mathrm{eV}}\xspace}

\newcommand*{\T}{\ensuremath{\,\mathrm{T}}\xspace}
\newcommand*{\mT}{\ensuremath{\,\mathrm{mT}}\xspace}
\newcommand*{\uT}{\ensuremath{\,\mu\mathrm{T}}\xspace}

\newcommand*{\dBm}{\ensuremath{\,\mathrm{dBm}}\xspace}

\newcommand*{\ms}{\ensuremath{\,\mathrm{ms}}\xspace}
\newcommand*{\s}{\ensuremath{\,\mathrm{s}}\xspace}
\newcommand*{\us}{\ensuremath{\,\mathrm{\mu s}}\xspace}
\newcommand*{\ns}{\ensuremath{\,\mathrm{ns}}\xspace}
\newcommand*{\rpm}{\ensuremath{\,\mathrm{rpm}}\xspace}
\newcommand*{\minute}{\ensuremath{\,\mathrm{min}}\xspace}
\newcommand*{\degree}{\ensuremath{\,^\circ\mathrm{C}}\xspace}

\renewcommand{\thefigure}{\arabic{figure}}
\newcommand*{\EqRef}[1]{Eq.\,(\ref{#1})}
\newcommand*{\FigRef}[1]{Fig.\,\ref{#1}}
\newcommand*{\TabRef}[1]{Tab.\,\ref{#1}}
\newcommand*{\dd}[2]{\mathrm{d}#1/\mathrm{d}#2}
\newcommand*{\ddf}[2]{\frac{\mathrm{\partial}#1}{\mathrm{\partial}#2}}
\newcommand*{\dddf}[2]{\frac{\mathrm{d}#1}{\mathrm{d}#2}}

\title{Coherence and Decay of Higher Energy Levels of a Superconducting Transmon Qubit}


 \author{Michael J. Peterer$^{1,2}$}
 \email{michael.peterer@physics.ox.ac.uk}
 \author{Samuel J. Bader$^{1}$}
 \author{Xiaoyue Jin$^1$}
 \author{Fei Yan$^{1}$}
 \author{Archana Kamal$^{1}$}
\author{Theodore J. Gudmundsen$^{3}$}
 \author{Peter J. Leek$^{2}$}
 \author{Terry P. Orlando$^{1}$}
 \author{William D. Oliver$^{1,3}$}
 \author{Simon Gustavsson$^{1}$}
 \affiliation{
$^{1}$Research Laboratory of Electronics, Massachusetts Institute of Technology, Cambridge, MA 02139, USA \\
$^{2}$Clarendon Laboratory, Department of Physics, University of Oxford, OX1 3PU, Oxford, United Kindom \\
 $^{3}$MIT Lincoln Laboratory, 244 Wood Street, Lexington, MA 02420, USA
}

\date{\today}


\begin{abstract} 
We present measurements of coherence and successive decay dynamics of higher energy levels of a superconducting transmon qubit. 
By applying consecutive $\pi$-pulses for each sequential transition frequency, we excite the qubit from the ground state up to its fourth excited level and characterize the decay and coherence of each state.
We find the decay to proceed mainly sequentially, with relaxation times in excess of 20$\us$ for all transitions. We also provide a direct measurement of the charge dispersion of these levels by analyzing beating patterns in Ramsey fringes. 
The results demonstrate the feasibility of using higher levels in transmon qubits for encoding quantum information.

\end{abstract}


\maketitle



\begin{figure*}[t!]
\centering
\includegraphics[width=\linewidth]{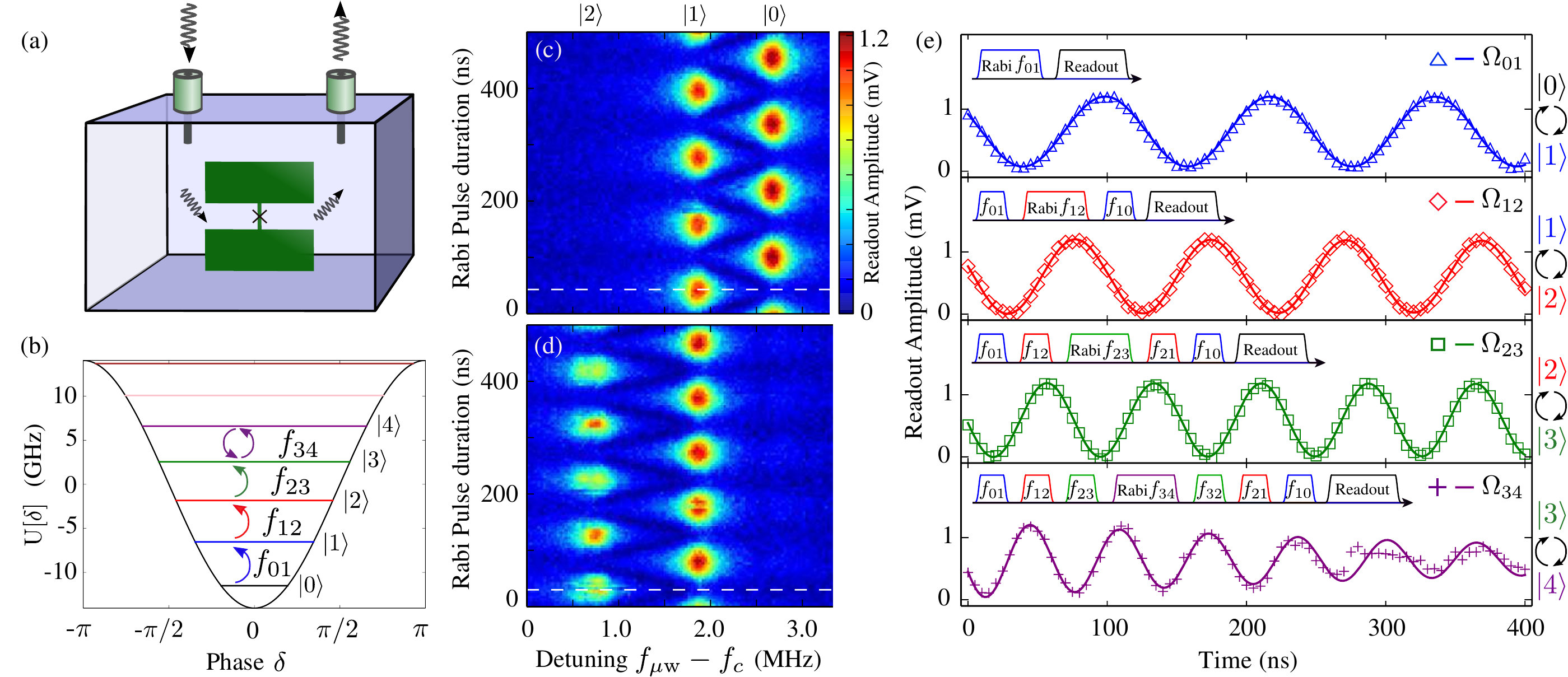}
\caption{(a) Schematic of the physical transmon qubit (not to scale) housed in the 3D cavity. The cross respresents the Josephson junction, situated between the two junction electrodes forming the capacitor. (b) Simulated energy spectrum of the transmon with parameters $E_{J}/E_{C} = 58$, where $U[\delta]$ is the Josephson potential. (c) Rabi oscillations between state $\ket{0}$ and $\ket{1}$ under a Rabi drive tone of varying duration at $f_{01}$. The white dashed line indicates the position of the first $\pi$-pulse at $\pi_{01} = 40 \ns$. (d) Rabi oscillations between $\ket{1}$ and $\ket{2}$ with $\pi_{12} = 29 \ns$, obtained by adding a Rabi drive tone at $f_{12}$ after initializing state $\ket{1}$ by applying $\pi_{01}$-pulse. (e) Rabi oscillations on each successive qubit transition up to state $\ket{4}$ using the depopulation readout method. The corresponding excitation pulse sequence and respective depopulation sequence are shown for each Rabi drive. The solid lines are best-fit curves allowing the extraction of the Rabi frequencies $\Omega_{ij}$ as $\Omega_{01} = 8.45 \MHz$, $\Omega_{12} = 10.3 \MHz$, $\Omega_{23} = 13.0 \MHz$, and $\Omega_{34} = 15.6 \MHz$ respectively.}
\label{fig:Rabi}
\end{figure*}

Universal quantum information processing is typically formulated with two-level quantum systems, or qubits \cite{Nielson:2010}. However, extending the dimension of the Hilbert space to a $d$-level system, or ``qudit,'' can provide significant computational advantage. In particular, qudits have been shown to reduce resource requirements \cite{Muthukrishnan:2000, Bullock:2005}, improve the efficiency of certain quantum cryptanalytic protocols \cite{Bechmann:2000,Cerf:2002,Durt:2004, Bregman:2008}, simplify the implementation of quantum gates \cite{Lanyon:2008, Chow:2013}, and have been used for simulating multi-dimensional quantum-mechanical systems \cite{Neeley:2009}.
The superconducting transmon qubit \cite{Koch:2007} is a quantum $LC$-oscillator with the inductor replaced by a Josephson junction [\FigRef{fig:Rabi}(a)]. The non-linearity of the Josephson inductance renders the oscillator weakly anharmonic, which allows selective addressing of the individual energy transitions and thus makes the device well-suited for investigating multi-level quantum systems.
The transmon's energy potential is shallower than the parabolic potential of an harmonic oscillator, leading to energy levels that become more closely-spaced as energy increases [\FigRef{fig:Rabi}(b)].  Although leakage to these levels can be a complication when operating the device as a two-level system \cite{Chow:2010}, the existence of higher levels has proven useful for implementing certain quantum gates \cite{DiCarlo:2009, Abdumalikov:2013}. Full quantum state tomography of a transmon operated as a three-level qutrit has also been demonstrated \cite{Bianchetti:2010}. 

In this work, we investigate the energy decay and the  phase coherence of the first five energy levels of a transmon qubit embedded in a three-dimensional cavity \cite{Paik:2011}. 
We find the energy decay of the excited states to be predominantly sequential, with non-sequential decay rates suppressed by two orders of magnitude.  The suppression is a direct consequence of the parity of the wave functions, in analogy with the orbital selection rules governing transitions in natural atoms.
We find that the sequential decay rates scale as $i$, where $i = 1,...,4$ is the initial excited state, thus confirming the radiation scaling expected for harmonic oscillators \cite{Lu:1989, Wang:2008}. The decay times remain in excess of $20\us$ for all states up to $i=4$, making them promising resources for quantum information processing applications. 
In addition, we characterize the quantum phase coherence of the higher levels by performing Ramsey-type interference experiments on each of the allowed transitions, and find strong beating in the resulting interference pattern, due to quasiparticle tunneling. This experimental result provides a direct measurement of the charge dispersion of the different levels \cite{Schreier:2008, Houck:2008,Catelani:2012,Sun:2012,Wang:2014,Riste:2013}.

Our device is a transmon qubit with a transition frequency $f_{01}=4.97\GHz$ for the first excited state, embedded in an aluminum 3D cavity with a bare fundamental mode $f_{c} = 11 \GHz$, and thermally anchored at a base temperature of $15\mK$ inside a dilution refrigerator. 
The interactions between the qubit in state $\ket{i}$ and the cavity causes a dispersive shift $\chi_i$ of the cavity resonance to a new frequency $f_i = f_c + \chi_i$, which is exploited for the readout of the qubit state \cite{Wallraff:2005}. We probe the state by sending coherent readout microwaves of frequency $f_{\mu \mathrm{w}}$ through the resonator at a chosen detuning $\Delta_{\mu \mathrm{w}} = f_{\mu\mathrm{w}} - f_c$ from the bare cavity resonance, and measure the averaged transmission coefficient $S_{21}$ of the signal over many experiments. Through a heterodyne detection scheme, the voltage amplitude of the transmission signal at $f_{\mu\mathrm{w}}$ is recorded, from which the qubit state occupation is then directly extracted. The resonator transmission takes the form of a Lorentzian peak $S_{21}^i (f_{\mu \mathrm{w}})  = p_i /[1+2\, \mathfrak{i}\, Q_t (f_{\mu\mathrm{w}}-f_i)/f_i]$ (see \cite{SM}), centered around the qubit state-dependent frequency $f_i$, with magnitude $p_i$ representing the state population, and $Q_t$ the total quality factor. When the total population $p$ is distributed over several states $\ket{i}$, the transmission becomes $S_{21}(f_{\mu \mathrm{w}}) = \sum_{i}{} S_{21}^i (f_{\mu \mathrm{w}})$. 


Exciting the transmon to a higher level first requires us to measure and analyse Rabi oscillations between adjacent pairs of energy levels, working sequentially up the ladder of states, as depicted in \FigRef{fig:Rabi}(b). Combined with qubit spectroscopy at each step, this protocol allows us to obtain the successive transition frequencies up to $f_{i-1,i}$ and to accurately calibrate the corresponding $\pi$-pulses.
Starting with the qubit in the ground state $\ket{0}$, we apply a microwave pulse at $f_{01}$ which drives the population between states $\ket{0} \leftrightarrow \ket{1}$ [see \FigRef{fig:Rabi}(c)].  As the qubit undergoes Rabi oscillations, the resonator transmission peak continuously rises and falls, oscillating between the discrete shifted resonance frequencies $f_0$ and $f_{1}$. Fitting the Rabi oscillations on state $\ket{1}$ permits us to experimentally extract the $\pi$-pulse duration $\pi_{01} = 40 \ns$ from the white dashed line in \FigRef{fig:Rabi}(c), required to achieve a complete population transfer at transition frequency $f_{01}$. In the second step, we add a second Rabi drive tone at $f_{12}$ promptly after the $\pi_{01}$-pulse (with a delay of $70\ns$, much shorter than the decay time $\Gamma_{10}^{-1}$ from state 1 to 0), so as to perform Rabi oscillations between states $\ket{1} \leftrightarrow \ket{2}$, enabling the calibration of the second $\pi$-pulse of duration $\pi_{12} = 29 \ns$ to reach $\ket{2}$. This process is repeated by adding a drive tone at each subsequent transition in order to calibrate the $\pi$-pulses up to the desired state. These procedures also allow us to experimentally extract the dispersive shifts $\chi_i$. A full numerical simulation of our coupled qubit-cavity Hamiltonian predicts all the qubit transition frequencies $f_{i-1,i}$ and the dispersive shifts $\chi_i$, and they are in very good agreement with the experimentally obtained values, displayed in \TabRef{tab:values}.
%


%
When driving Rabi oscillations on the transition $\ket{i} \leftrightarrow \ket{i+1}$ for $i\geq 2$, the readout by the method presented above is not possible in this device, because state $\ket{3}$ does not appear as a conditional shift to the resonator. This is a consequence of the fact that certain states escape the dispersive regime due to their mixing with higher-excited states that have transition frequencies close to the resonator frequency, see simulation in \cite{SM}.
As a result, we use a modified readout protocol, which does not require measurement pulses at the shifted resonance $f_{3}$ or $f_{4}$. After preparing the qubit in state $\ket{i}$ via the upward sequence of $\pi$-pulses $S_i^{\uparrow} = (\pi_{01}, \pi_{12}, \dotsc, \pi_{i-1,i})$, we additionally apply a depopulation sequence $S_i^{\downarrow} = (\pi_{i,i-1}, \dotsc, \pi_{21}, \pi_{10})$ to the qubit immediately before the readout. This maps the population $p_i$ of state $\ket{i}$ onto that of the ground state $\ket{0}$, allowing us to measure $p_i$ by simply probing the resonator at the frequency $f_0$. 

By incorporating the depopulation sequence we are able to drive Rabi oscillations of the transmon up to state $\ket{4}$, as shown in \FigRef{fig:Rabi}(e). The Rabi frequencies $\Omega_{ij}$, extracted via a best-fit curve, are proportional to the matrix elements $\bra{i} \hat{n} \ket{j}$ between the states $i$ and $j$, where $\hat{n}$ denotes the number of Cooper pairs transferred between the two junction electrodes forming the capacitor \cite{SM}. Consequently, $\Omega_{ij}$ increase as $\bra{i} \hat{n} \ket{j} \propto \sqrt{j} $, as expected from the coupling between the transmon states and the resonator \cite{Koch:2007}.
Having thus obtained all the transition frequencies and $\pi$-pulse calibrations, the qubit can be initialized in any state up to $\ket{4}$ with the sequence $S_4^{\uparrow}$, and we proceed to investigate the decay and phase coherence of these higher levels.

\begin{table}[b!]
\begin{ruledtabular}
\begin{tabular}{l |l| l| l| l}
	Frequency & $f_{01}^*$ & $f_{12}^*$ & $f_{23}$ & $f_{34}$\\
	\hline
	Exp. $f$ ($\GHz$) &  4.9692 & 4.6944 & 4.3855 & 4.0280 \\
        Sim. $f$ ($\GHz$) &  $4.9692^*$ & $4.6944^*$ & 4.3874 & 4.0475\\
	\hline
	\hline
	Sequ. decay$^{-1}$ &  $\Gamma_{10}$ & $\Gamma_{21}$ & $\Gamma_{32}$ & $\Gamma_{43}$  \\
	\hline
	 time ($\us$)  & 84 $\pm$ 0.24 & 41 $\pm$ 0.21 & 30 $\pm$ 0.21 & 22 $\pm$ 2\\
       \hline
       \hline
	Non-sequ.$^{-1}$&  $\Gamma_{20}$ & $\Gamma_{31}$ & $\Gamma_{30}$ &  \\
	\hline
	time ($\us$)  & 1812 $\pm$ 223 & 1314 $\pm$ 359 & 2631 $\pm$ 694&   \\
	\hline
       \hline
       Dephasing $T_2$ &  $T_{2 \,(01)}$ & $T_{2 \,(12)}$ & $T_{2 \,(23)}$ & $T_{2 \,(34)}$ \\
	\hline
	time ($\us$) $\pm 20\%$  & 72 & 32  & 12 & $<$2  \\
\end{tabular}
\bigskip	
\begin{tabular}{l |l l l l l l l}
        
	Qubit State $i$ & $\ket{0}$ & $\ket{1}$ & $\ket{2}$ & $\ket{3}$ & $\ket{4}$\\
        \hline
	Exp. $\chi_{i}$  ($\MHz$) & 2.8&2&0.88& &\\
	Sim. $\chi_{i}$ ($\MHz$) & $2.8^*$&2&0.85&&\\
	Exp. $\epsilon_{ij}$ ($\MHz$) & & - & 0.09  & 2.53  & 5-10 \\
	Sim. $\epsilon_{ij}^{(max)}$ ($\MHz$) & & 0.0025  & 0.091  & 1.89  & 26.8 \\
\end{tabular}
\end{ruledtabular}
\caption{Comparison of experimental and simulated values for the transition frequencies $f_{i,i+1}$, the relaxation times $\Gamma^{-1}_{ij}$ for the sequential and non-sequential rates, the dephasing times $T_{2\,(ij)}$ for the superpositions of states $\ket{i}$ and $\ket{j}$, and the dispersive shifts $\chi_i$. The measured charge dispersion splittings $\epsilon_{ij}(n_g)$ extracted from Ramsey fringes are compared to the simulated maximum splittings $\epsilon_{ij}^{(max)}$. The asterisks indicate the values that were fitted to the experiment for use as parameters in the full numerical simulation of the coupled qubit-cavity Hamiltonian.}
\label{tab:values}
\end{table}

\begin{figure}[t!]
\centering
\includegraphics[width=\linewidth]{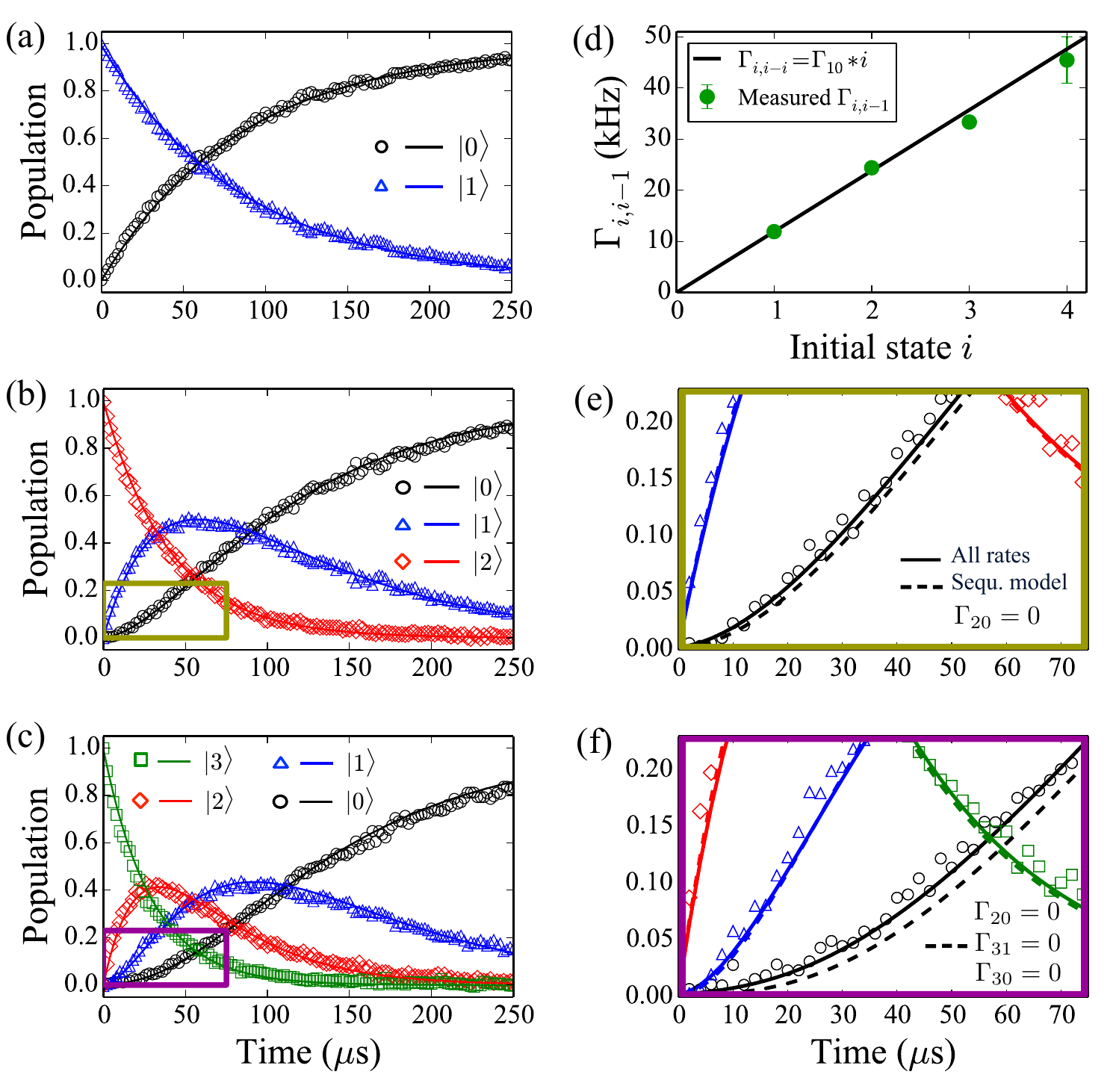}
\caption{(a)-(c) Population decay traces of the qubit states up to $\ket{3}$, obtained by varying the time delay $\Delta t$ before the depopulation sequence. The solid lines are state occupations from the multi-level decay model taking into account all decay channels. (d) The sequential decay rates $\Gamma_{i,i-1}$ (green dots) for increasing energy state $i$, showing the roughly linear dependance (solid line). (e)(f) Zoom of decay curves in panel (b) and (c) respectively, showing the model with all transition rates allowed (solid line) compared to the model with only neighboring transitions allowed (dashed line). The extracted decay times are listed in \TabRef{tab:values}.}
\label{fig:Decay vs State}
\end{figure}

We start by measuring the dynamics of the state population decay by introducing a varying time delay before the readout process. The calibrated and normalized population evolutions starting from states $\ket{1}$, $\ket{2}$, and $\ket{3}$ are plotted in \FigRef{fig:Decay vs State}(a)-(c). The decay from state $\ket{4}$ has also been measured and is presented in \cite{SM}.
%
We model the data with a multi-level rate equation describing the evolution of the state population vector $\vec{p}$, with $\Gamma_{ij}$ representing the decay rate from state $i$ to state $j$:
$
 \dd{\vec{p}(t)}{t}= \Gamma^{\sf T} \cdot \vec{p}(t).
$
The decay rates matrix $\Gamma$ has diagonal elements $(\Gamma)_{jj} = - \sum_{k=0}^{j-1}  \Gamma_{jk}$ and off-diagonal elements $\Gamma_{ij}$ for $i\neq j$. The upward rates are considered to be negligible by setting $\Gamma_{ij} = 0$ for all $i<j$, as $k_B T \ll h f_{ij}$ for all $i,j$. Indeed, the quiescent state-$\ket{1}$ population of our transmon is measured to be less than 0.1\% \cite{Jin:2014}. 
The state occupations of the model are plotted (solid lines) and compared to the experimental data in \FigRef{fig:Decay vs State} for each state, whereby the rates $\Gamma_{ij}$ are used as fitting parameters to extract all the system's relaxation rates. The fitting was performed iteratively, starting with the decay from $\ket{1}$, where we fit $\Gamma_{10}$ and then fix it for the next decay from $\ket{2}$, where $\Gamma_{21}$ and $\Gamma_{20}$ are determined, and so forth. 

The most prominent feature of the data is that the decay proceeds mainly sequentially \cite{Dykman:1984}, with the non-sequential decay rates suppressed by two orders of magnitude.  The extracted decay times are in excess of 20$\us$ for all states up to $\ket{4}$, and are listed in \TabRef{tab:values}. For the sequential rates, we find that the rates scale linearly with state $i$, as plotted in \FigRef{fig:Decay vs State}(d);  this behavior is consistent with decay processes related to fluctuations of the electric field (like Purcell or dielectric losses), for which we expect the lifetimes to be inversely proportional to $\vert\bra{i} \hat{n} \ket{j}\vert^2$ (see \cite{SM} for numerical calculations of the matrix elements).
 Furthermore, theoretical relaxation rates between neighboring levels due to quasiparticle tunneling also respect this approximate dependance $\Gamma_{i,i-1} \simeq i \,\Gamma_{10}$ \cite{Catelani:2011, Catelani:2011PRB}.
We note that the anharmonicity of this device is sufficiently weak that its decay rates scale as those of Fock states in a harmonic oscillator \cite{Lu:1989, Wang:2008}.

\begin{figure}[t!]
\centering
\includegraphics[width=\linewidth]{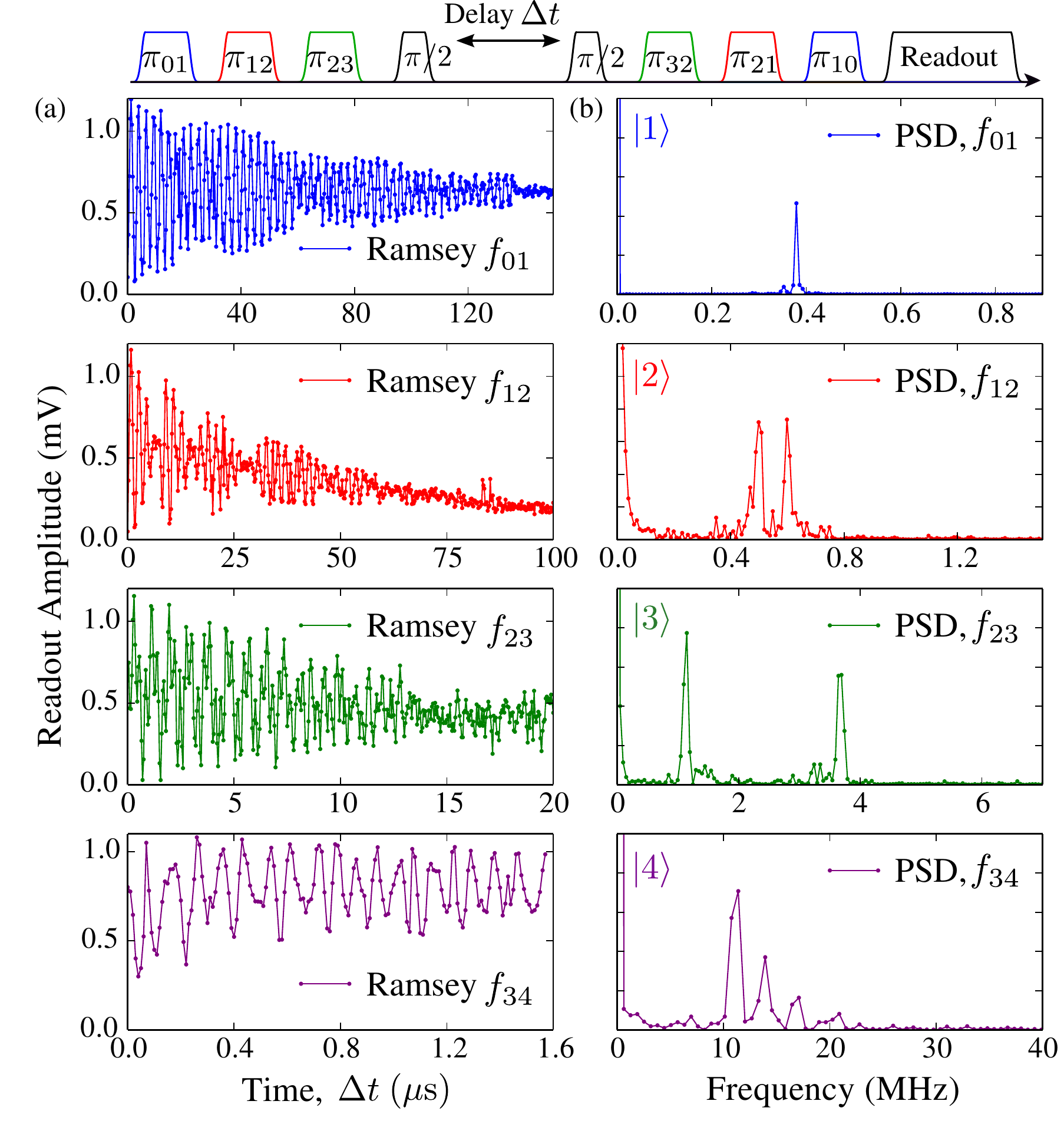}
\caption{(a) Ramsey oscillations experiment on each subsequent energy transition up to state $\ket{4}$. (b) The Power Spectral Density (PSD), obtained from the discrete Fourier transforms of the corresponding Ramsey fringes.
The Ramsey pulse sequence (top) corresponds to the fourth-row panel, representing a Ramsey sequence on state $\ket{4}$, with the black $\pi/2$-pulses representing the $\pi_{34}/2$-pulses performed at frequency $f_{34}$ to bring the transmon into the superposition state $(\ket{3} + \ket{4})/\sqrt{2}$ before allowing the free evolution time $\Delta t$.}
\label{fig:Ramsey vs State}
\end{figure}

To illustrate the effect of the non-sequential rates, we also fitted the data to a model involving only sequential rates (dashed lines in \FigRef{fig:Decay vs State}(e)(f)). Although the deviations between the two fits are small, inclusion of the rates $\Gamma_{20}^{-1}$,$\Gamma_{30}^{-1}$ and $\Gamma_{31}^{-1}$ does provide somewhat better matching for the initial increase in ground state population $\ket{0}$ for $t<70\us$ where we would expect the largest impact.  
From numerical simulations of the qubit-resonator Hamiltonian, we expect the rates $\Gamma_{20}$ and $\Gamma_{31}$ to be strongly suppressed due to the parity of those states \cite{Deppe:2008}, whereas the matrix element $\abs{\bra{3} \hat{n} \ket{0}}^2$ relevant for $\Gamma_{30}$ is about 100 times smaller than $\abs{\bra{1} \hat{n} \ket{0}}^2$ \cite{SM}. Quasiparticles also contribute to relaxation rates for non-neighboring levels, and theory \cite{Catelani:2012} predicts that they are suppressed by at least three orders of magnitude, a much stronger suppression than we extract. This suggests that the non-sequential decay rates are dominated by some non-quasiparticle process, such as dielectric loss or coupling to other cavity modes.

We now proceed to investigate the phase coherence of the higher levels by performing a Ramsey-fringe measurement, whereby we obtain the dephasing times $T_2$.
A Ramsey experiment on state $\ket{i}$ consists of first applying $\pi$-pulses to bring the transmon to state $i-1$, followed by a $\pi/2$-pulse at frequency $f_{i-1,i}$ to bring it into a superposition of states $\ket{i-1}$ and $\ket{i}$, then allowing a variable free-evolution time $\Delta t$ to pass, and finally applying a second $\pi/2$-pulse before applying the depopulation sequence and performing the readout.
The measured Ramsey fringes are shown in \FigRef{fig:Ramsey vs State} for each state up to $\ket{4}$.  The frequency of the $\pi/2$-pulses was purposefully detuned to generate oscillating traces.
%
%
The power spectral density of the data, obtained via a discrete Fourier transform, reveals two well-defined frequency components for states $\ket{2}$ and $\ket{3}$, and a number of frequencies for state $\ket{4}$. As described in the supplementary material, we fit the Ramsey fringes in \FigRef{fig:Ramsey vs State}(a) to a sum of two damped sinusoids, and the extracted dephasing times $T_2$ are listed in \TabRef{tab:values}.

\begin{figure}[b!]
  \centering
  \includegraphics[width=\linewidth]{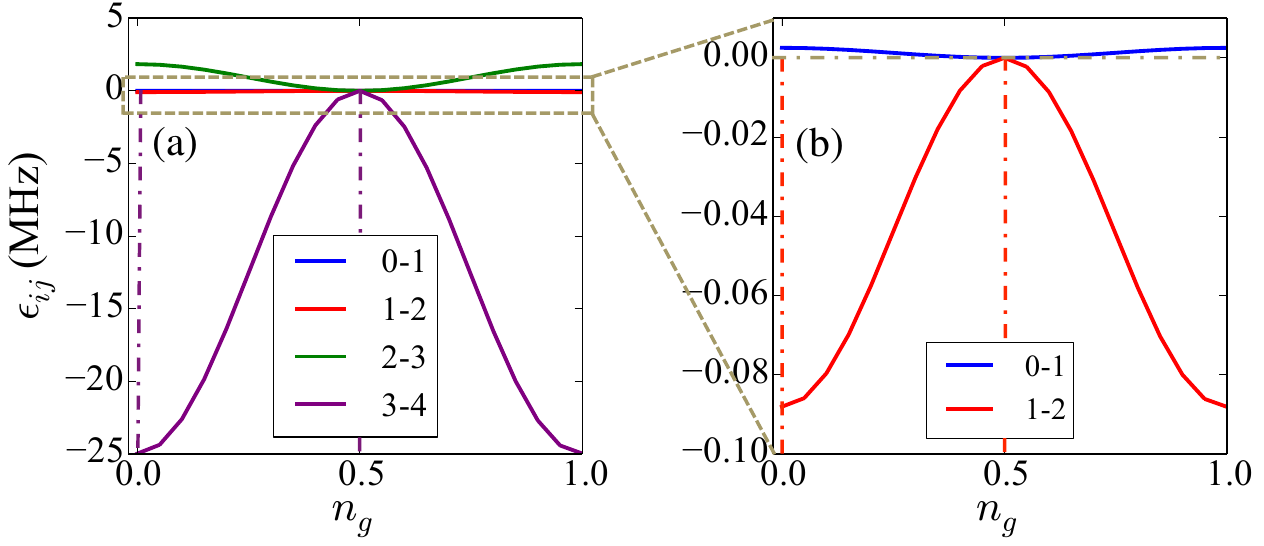}
  \caption{(a) Calculated change in transition frequency $\epsilon_{ij} (n_g)$  as a function of the effective offset charge $n_g$, expressing the charge dispersion of each transition. 
  (b) Zoom of the lowest two transitions 0-1 and 1-2.}
  \label{fig:dispersion}
\end{figure}

The splitting of the transition frequencies can be understood in terms of quasi-particle tunneling between the two junction electrodes \cite{SM}. Despite the large $E_J/E_C$ ratio, the transmon retains some sensitivity to charge fluctuations, and the charge dispersion approximately grows in an exponentially way with increasing level number \cite{SM}.  
From our full numerical transmon-cavity simulation, we calculate the change in level splitting $\epsilon_{ij}(n_g)$ between levels $i,j$ as a function of the effective offset charge $n_g$ \cite{Koch:2007}, shown in \FigRef{fig:dispersion}. The maximum change in $\epsilon_{ij}$  due to quasi-particle tunneling is given by $\epsilon_{ij}^{(max)} = \epsilon_{ij}(n_g\!=\!1/2) - \epsilon_{ij}(n_g\!=\!0)$, as marked by vertical dashed lines, but note that what we measure experimentally is the dispersion between $n_g +1/2$ and $n_g$ for an unknown value of $n_g$.
The measured splittings are compared to the calculated maximum splittings $\epsilon_{ij}^{(max)}$ in \TabRef{tab:values}. State $\ket{1}$ is unresolved, in agreement with the small splitting of 2.5 $\kHz$ predicted, whereas we find that the splittings of states $\ket{2}$ and $\ket{3}$ are reasonably well predicted by the simulation.  Charge traps between the substrate and the deposited metal film, or the presence of two-level fluctuators in the junction, also lead to charge fluctuations, possibly explaining the additional peaks seen in the spectrum of the state $\ket{4}$.
It should be noted that for quantum information purposes, the noise causing the beating in the Ramsey fringes can be refocused with an echo sequence by adding a temporally short $\pi$-pulse (broad frequency spectrum) to the center of the Ramsey sequence \cite{Hahn:1950}.

 
In conclusion, we have demonstrated the preparation and control of the five-lowest states of a transmon qubit in a three-dimensional cavity.  We observed predominantly sequential energy relaxation, with non-sequential rates suppressed by two orders of magnitude. In addition, our direct measurement of the charge dispersion at higher levels agrees well with theory and facilitates further studies of the crucially important dephasing characteristics of quantum circuits. The measured qubit lifetimes in excess of 20$\us$ at energy states up to $\ket{4}$ expands the practicability of transmons for quantum information applications and simulations using multi-level systems. 

We thank G. Catelani, J. Bylander, A. P. Sears, D. Hover, J. Yoder, and A. J. Kerman for helpful discussions, as well as Rick Slattery for assistance with the cavity design and fabrication, and George Fitch for assistance with layout. This research was funded in part by the Assistant Secretary of Defense for Research \& Engineering under Air Force Contract FA8721-05-C-0002, by the U.S. Army Research Office (W911NF-12-1-0036), and by the National Science Foundation (PHY-1005373). MJP and PJL acknowledge funding from the UK Engineering and Physical Sciences Research Council. 

\bibliographystyle{apsrev4-1}
\bibliography{HigherLevels} 

\clearpage


\setcounter{figure}{0}
\setcounter{table}{0}

\renewcommand{\thetable}{S\arabic{table}}   
\renewcommand{\thefigure}{S\arabic{figure}}

\begin{widetext}
\section{\fontsize{13}{13} \selectfont S\lowercase{upplementary} M\lowercase{aterial} \lowercase{to} 
 "C\lowercase{oherence} \lowercase{and} D\lowercase{ecay} \lowercase{of} H\lowercase{igher} E\lowercase{nergy} L\lowercase{evels} \\
\lowercase{of a} S\lowercase{uperconducting} T\lowercase{ransmon} Q\lowercase{ubit}"}
\vspace{0.5cm}


In the following Supplementary Material we present a detailed description of the measurement techniques, the data analysis, and further decay data to Fig.2. Furthermore, we present the full numerical simulation of the transmon and the coupled qubit-cavity Hamiltonian.

\vspace{0.5cm}

\end{widetext}

\section{Device characterization}

Our superconducting transmon qubit consists of a single nanofrabricated Josephson junction (Al/Al$\mathrm{O}_x$/Al) contacted between two large electrodes of sizes $600\um$ x $250\um$ that form the capacitor. This circuit is fabricated on a $5\mm$ x $5\mm$ sapphire chip, which is embedded in an aluminum 3D cavity with a bare fundamental mode $f_{c} = 10.97537 \GHz$, thermally anchored at a base temperature of 15$\mK$ inside a dilution refrigerator. Two SMA coupled ports allow microwave signals in and out of the cavity at a rate of $\kappa/2\pi \approx  100 \kHz$. By comparing the measured dispersive shifts of the resonator with the numerical simulation, we determined the Josephson energy $E_J = 14.07 \GHz$ and charging energy $E_C = 243 \MHz$ with a ratio $E_{J}/E_{C} = 58$, placing it in the charge-insensitive transmon regime. It has a non-tunable first transition frequency $f_{01} = 4.9692 \GHz$, a detuning from the cavity $\Delta = |f_c - f_{01}| = 6.00617 \GHz$, and a coupling strength $g_{0}/2\pi = 164.5 \MHz$. The measured relaxation times and dephasing times are listed in Tab\,I. During a different cooldown, the following additional values for our transmon were measured : $T_{2}^* = 90 - 115 \us$, and with spin echo $T_{2E} = 154 \us$. 

In our experiments, all microwave pulse sequences to control the qubit up to state $\ket{4}$ are generated via single-sideband mixing (upconversion by an I/Q mixer) from a 12GS/s Tektronix AWG 7122 and a carrier signal of frequency 3.5\GHz from a \textit{single} Agilent 8267D PSG vector signal generator. The AWG's analog bandwidth of 3.2\GHz is sufficiently large, and the anharmonicity of the transmon sufficiently weak, to allow the upconversion of microwave pulses into a range of frequencies large enough to access all transition frequencies from $f_{01}$ to $f_{34}$ without the need for numerous signal generators. From the measured transition frequencies listed in Tab.\,I, the anharmonicities $\alpha_{ij} = f_{01}-f_{ij}$ are found to be $\alpha_{12} = 274.8\MHz$,  $\alpha_{23} = 583.7\MHz$, and $\alpha_{34} = 941.2\MHz$ respectively, well within the AWG's bandwidth. 

All drive pulses have flat-top sections of variable duration and Gaussian-shaped rise and fall envelopes with fixed duration of $20\ns$, chosen to prevent undesired leakage to neighboring levels. The delay between pulses in a drive sequence is constant and set to $70\ns$, which is much shorter than any of the sequential decay times.

\begin{figure}[h!]
\centering
\includegraphics[width=\linewidth]{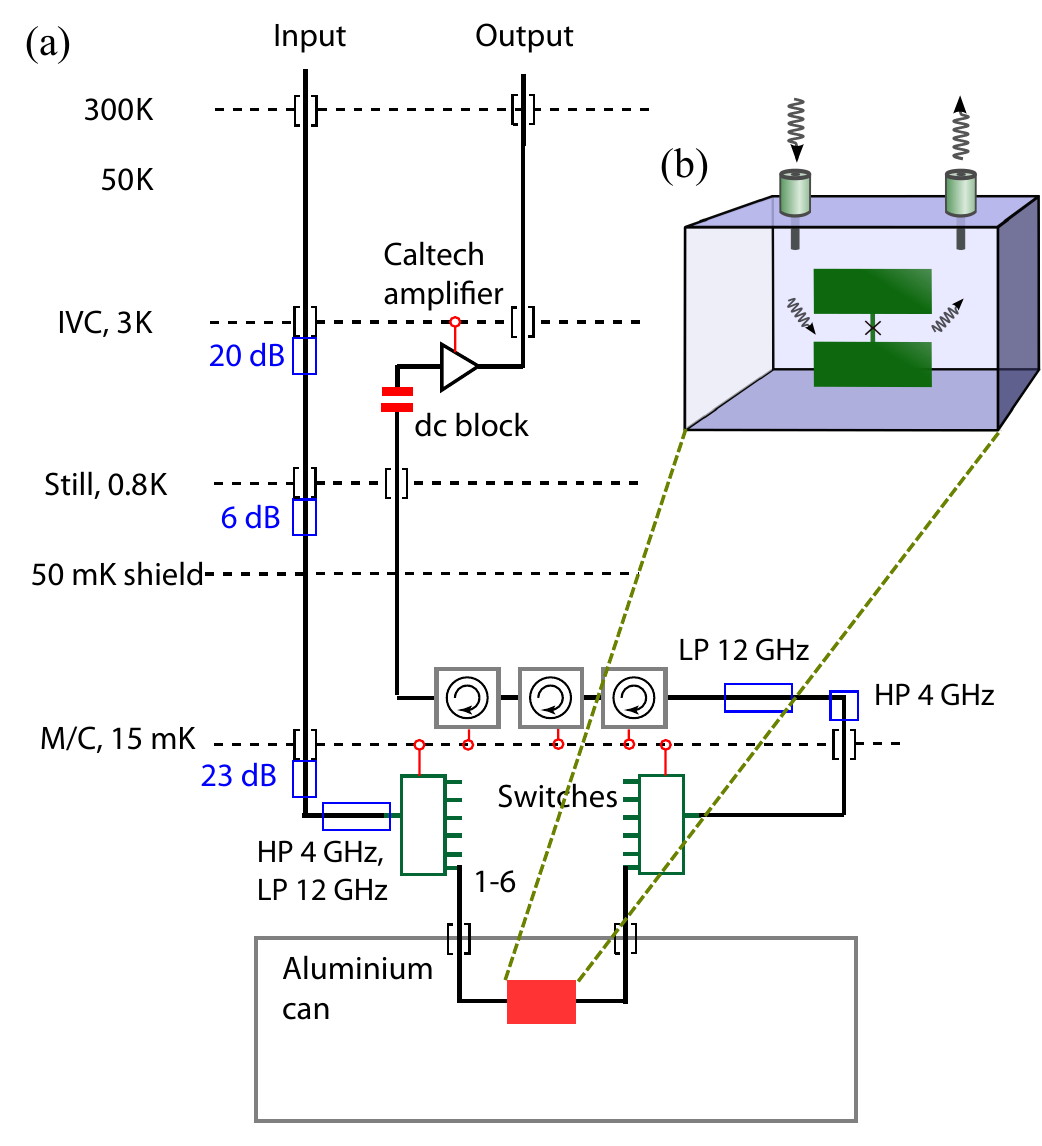}
\caption{(a) Schematic of the measurement setup, showing the microwave circuitry and temperature stages of the dilution refrigerator. (b) Schematic of the transmon qubit (not to scale) embedded in the three-dimensional cavity.}
\label{fig:Setup}
\end{figure}

\section{Extraction of the state populations}

\begin{figure}[t!]
\centering
\includegraphics[width=\linewidth]{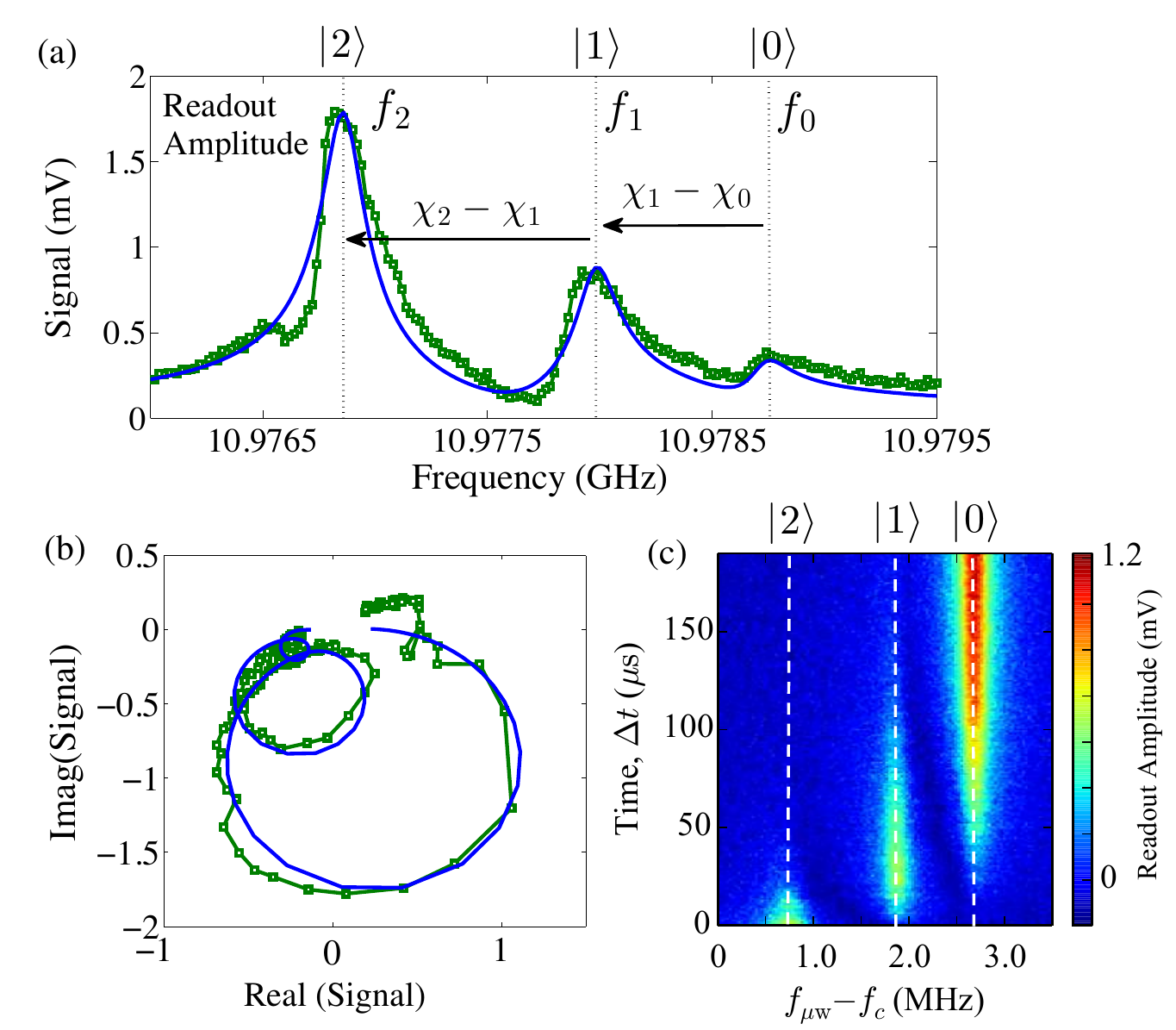}
\caption{(a) Transmission spectrum of the measured trace (green markers) at readout delay $t_R=8\us$ from the population decay plot in (c) of an initially prepared state $\ket{2}$. The acquired signal is the averaged readout amplitude of the cavity response and is proportional to the transmission coefficient $S_{21}$. The best-fit (blue) is a sum of overlapping Lorentzian peaks centered at frequencies $f_2$,$f_1$, and $f_0$, corresponding to the qubit being in state $\ket{2}$, $\ket{1}$, and $\ket{0}$ respectively, and expressing the dispersive shifts $\chi_1$ and $\chi_2$. (b) Transmission signal fitted in the complex plane in order to consider the phase in addition to the amplitude.
(c) Qubit population decay from level $\ket{2}$ to $\ket{1}$ to $\ket{0}$, obtained by varying the time delay $\Delta t$ before readout. The frequency sweep of the readout pulse detuning reveals the temporal transition of the resonator frequency from $f_2$ to $f_1$ to $f_0$.}
\label{fig:lorentzians}
\end{figure}

\begin{figure}[b!]
\centering
\includegraphics[width=\linewidth]{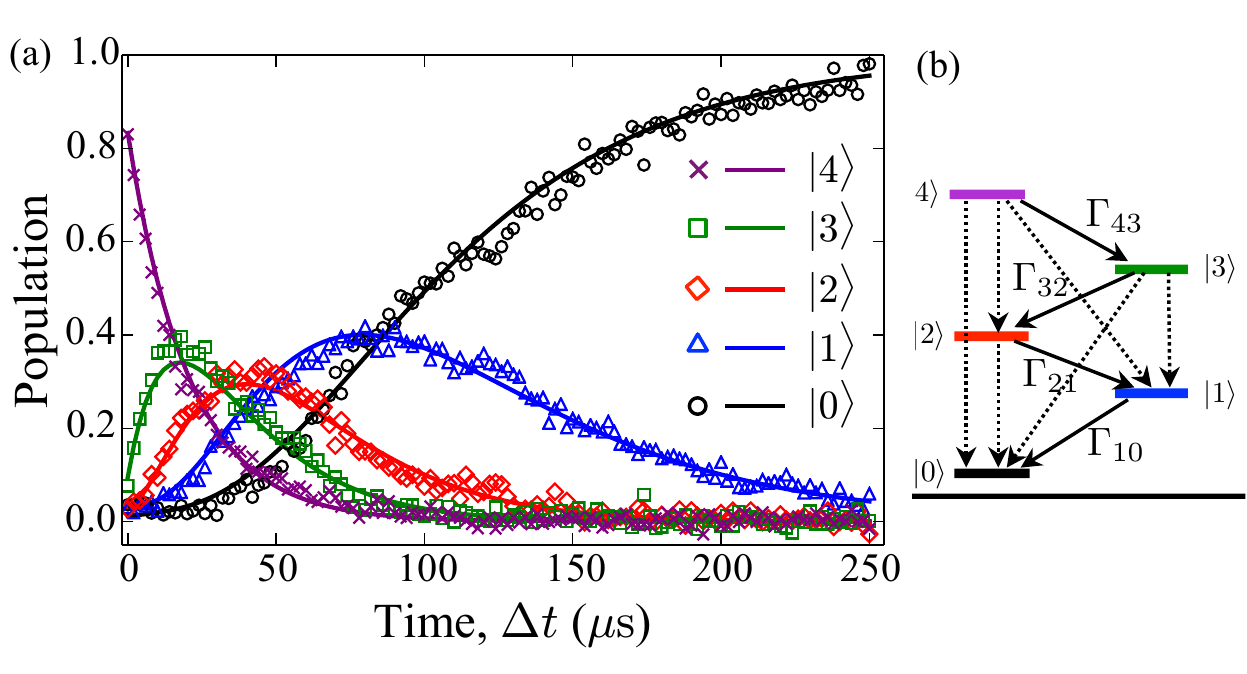}
\caption{(a) Measured decays of the qubit populations up to energy state $\ket{4}$. The data represents the populations after correction for overlapping Lorentzians and relaxation during readout. The solid lines are numerical solutions to the sequential decay model. (b) Energy level scheme showing the sequential decay rates via the solid arrows.}
\label{fig:Decay vs State4}
\end{figure}

In the following, we describe how the measured data of the population decays in \FigRef{fig:lorentzians}(c) and \FigRef{fig:Decay vs State4} must be corrected to obtain the actual normalized populations. The data from each decay trace is the measured voltage $V_i(t)$ detected by the data acquisition card for a readout pulse at $f_0$. However, the data for each trace must be corrected for the overlap of the Lorentzian tails from the other populations present in frequency space at the point $f_0$, as displayed in \FigRef{fig:lorentzians}(a), as well as for the decay during the readout.
We represent the evolution of the measured populations as $\vec{V}(t) = [V_0(t),V_1(t),V_2(t),V_3(t),V_4(t)]^{\sf T}$ and define the value at $f_k$ of an individual cavity response Lorentzian centered at $f_i$ as 
\begin{equation}
L_{ik} = L_i(f_k) = \frac{1}{ [1+2\, \mathfrak{i}\, Q_t (f_k-f_i)/f_i]},
\end{equation} 
with $Q_t$ the total quality factor. In our first decay experiment, the total population $p$ is distributed over several states $\ket{i}$ and the transmission spectrum becomes
$L_k(f_k) = \sum_{i}{} L_i(f_k)$.

The corrected state populations $p_{i}^*(t) \in \vec{p}^*(t)$ are obtained by $\vec{p}^*(t) = L^{-1} \cdot \vec{V}(t)$, with the inversion matrix given by
\begin{equation}
L = 
	\begin{pmatrix}
	L_{00} & L_{10} & L_{20} & L_{30} & L_{40} \\
	L_{10} & L_{00} & L_{20} & L_{30} & L_{40} \\
	L_{10} & L_{20} & L_{00} & L_{30} & L_{40} \\
	L_{10} & L_{20} & L_{30} & L_{00} & L_{40} \\
	L_{10} & L_{20} & L_{30} & L_{40} & L_{00} \\
	\end{pmatrix}.
\end{equation}
The matrix elements are obtained by fitting the Lorentzian transmission profile in the complex plane to the measured voltages, as shown in \FigRef{fig:lorentzians}(b) for a sample trace at delay time $t_R=8 \us$.

The data $\vec{p}^*(t)$ must then be corrected for the relaxation of the qubit during the readout time of $T_{read}= 8 \us$. The card averages the signal over the time of acquisition $T_{read}$, during which the population from the Lorentzian $p_0^*(t)$, mapped to center $f_1$ after the depopulation sequence (corresponding to qubit in $\ket{1}$), relaxes by a factor 
\begin{equation}
\bar{\Lambda} = \int_{T_{read}} \, \mathrm{e}^{-\Gamma_{10} t}\, \mathrm{d} t,
\end{equation}
where the rate $\Gamma_{10}$ is extracted from the decay model fit. The final true populations are therefore obtained by
\begin{equation}
\left\{ 
  \begin{array}{l l}
    p_i = p_i^* - p_0^* (\bar{\Lambda}^{-1} -1)& \quad \textrm{for each} \,\,i = 1,2,3,4 \\
    p_0 = p_0^* - p_1^* (\bar{\Lambda}^{-1} -1)  & \quad \\
  \end{array} \right.
\end{equation}
and are plotted in \FigRef{fig:Decay vs State4} after being normalized.

When performing readout with the depopulation method, these cumbersome steps for extracting the populations can be replaced by the simpler method of directly measuring calibration values for the population of each state. This is done by pumping the population to state $\ket{i}$ and measuring the readout voltage response at $f_0$. Then the inversion matrix $L_{cal}$ contains simply the calibration voltage values measured as
\begin{equation}
L_{cal} = 
	\begin{pmatrix}
	2.3 & 0& 0 & 0  \\
	0.4 & 2.3&0 & 0 \\
	0.4& 0.25 &2.3& 0  \\
	0.4 & 0.25 & 0.2 & 2.3  \\
	\end{pmatrix}.
\end{equation}
\vspace{0.01cm}

The calibrated populations $p_i(t) \in \vec{p}(t)$ are thus given by 
\begin{equation}\label{eq:inversion2}
\vec{p}(t) = L_{cal}^{-1} \cdot \vec{V}(t).
\end{equation}
This method also intrinsically calibrates for the decay during readout and is the method used to acquire the normalized population decay traces in Fig.\,2.

\section{Extraction of dephasing times}

\begin{figure}[t!]
\centering
\includegraphics[width=\linewidth]{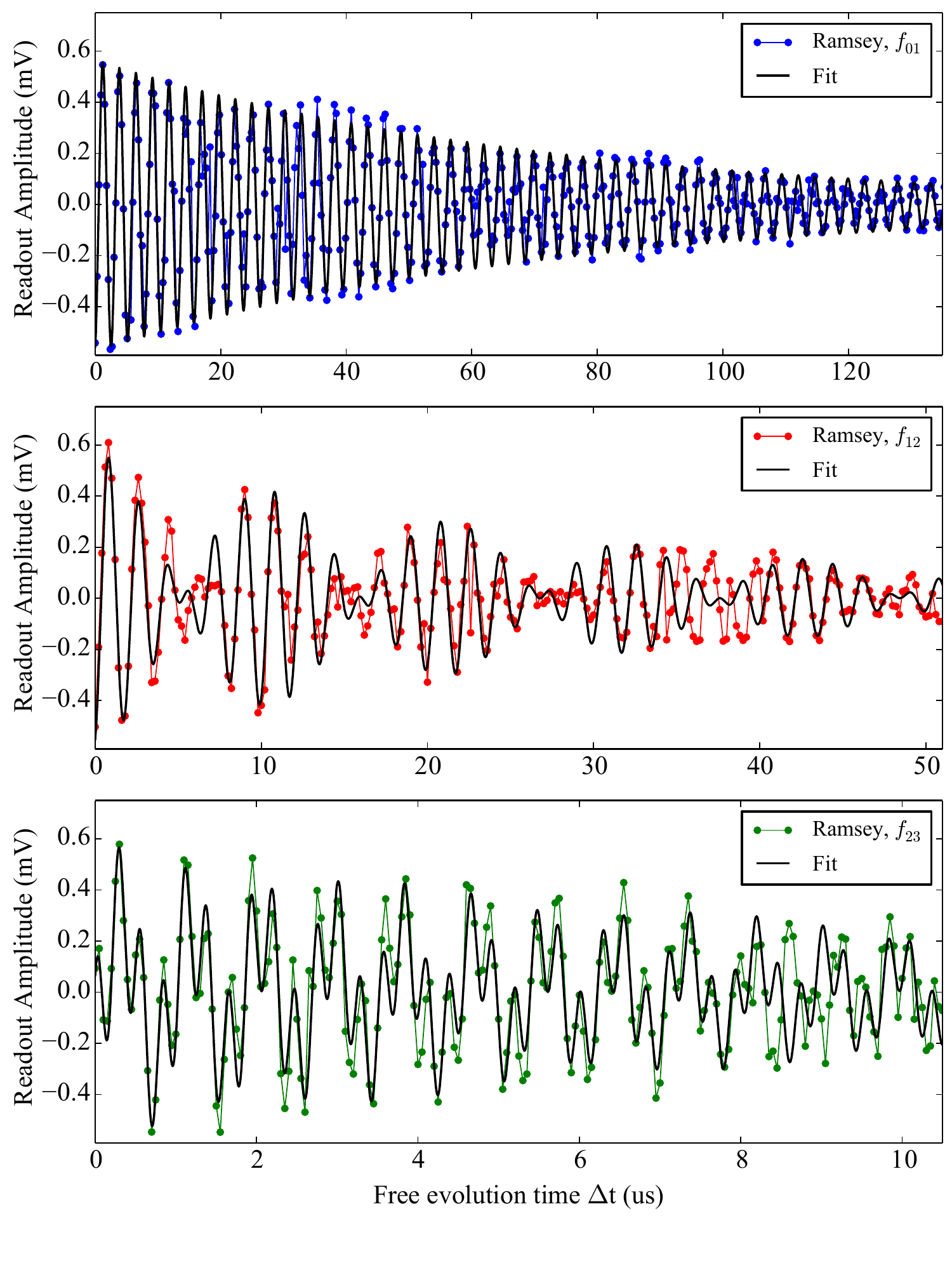}
\caption{(a) Fits of the Ramsey oscillations on each subsequent energy transition up to state $\ket{3}$. The Ramsey traces are plotted after substracting a smooth background to discard the energy decay towards state $\ket{0}$. The best fits reveal the dephasing times $T_{2\,(i,j)}$ for the superpositions between states $\ket{i}$ and $\ket{j}$, for $i,j = 1,2,3$, where each $T_2$ time is qualitatively defined as the decay time for the amplitude of the fringes. The measured Ramsey fringes on state $\ket{4}$ have a beating pattern containing several frequencies, making it impossible to obtain a reasonable fit.}
\label{fig:Ramsey fits}
\end{figure}

The measured Ramsey oscillations in Fig.\,3 directly reveal that the phase coherence for higher states become shorter.  The coherence times $T_{2\,(ij)}$ for the superpositions between states $\ket{i}$ and $\ket{j}$, for $i,j = 0,1,2,3,4$ \,\,($i<j$ and $j=i+1)$, are qualitatively defined as the decay time for the amplitude of the fringes. As explained in our Letter, the states $\ket{2}, \ket{3}$ and $\ket{4}$ show clear modulation in addition to the main oscillations, and leakage to lower levels causes the amplitude oscillations to drift toward zero. The discrete Fourier transform of the data (Fig.\,3b) reveals two well-defined frequency components for states $\ket{2}$ and $\ket{3}$, and a number of frequencies for state $\ket{4}$ (making it impossible to obtain a reasonable fit for this last state).  
In order to fit the Ramsey fringes for states up to $\ket{3}$, we first substract a smoothed background to discard the energy decay toward state $\ket{0}$. The voltage amplitude $A$ of the oscillations is then fitted to exponentially damped double sine curves
\begin{equation}\label{eq:Ramsey-fit}
A = e^{-t/T_{2}} [\cos(2\pi f_A t) + \cos(2\pi(f_A + \Delta f)t)],
\end{equation}
and displayed in \FigRef{fig:Ramsey fits}. The extracted $T_{2\,(ij)}$ values are listed in Tab.\,I, and are found to be acurate within 20\%. The frequencies $f_A$ and $f_A + \Delta f$ represent the two frequency components that fit the beating pattern in each Ramsey fringes plot, and have values $[f_A , \Delta f ]_{(ij)} = [379 \kHz, 0 \Hz]_{(01)}, [504\kHz, 93\kHz]_{(12)}$, and $[1.1\MHz, 2.5\MHz]_{(23)}$. The fit parameter $\Delta f$ represents the total charge dispersion splittings $\epsilon_{ij}(n_g) \leq \epsilon_{ij}^{(max)}$ for each transition between state $i$ and $j$, reported in Tab.\,I, with $j = i+1$ and $\epsilon_{ij} = \epsilon_{ij} (n_g\!=\!1/2) - \epsilon_{ij} (n_g\!=\!0)$.

\section{Simulation of the transmon and the coupled qubit-cavity Hamiltonian}
\subsection{Hamiltonians}

\begin{figure}[t!]
  \centering
  \includegraphics[width=\linewidth]{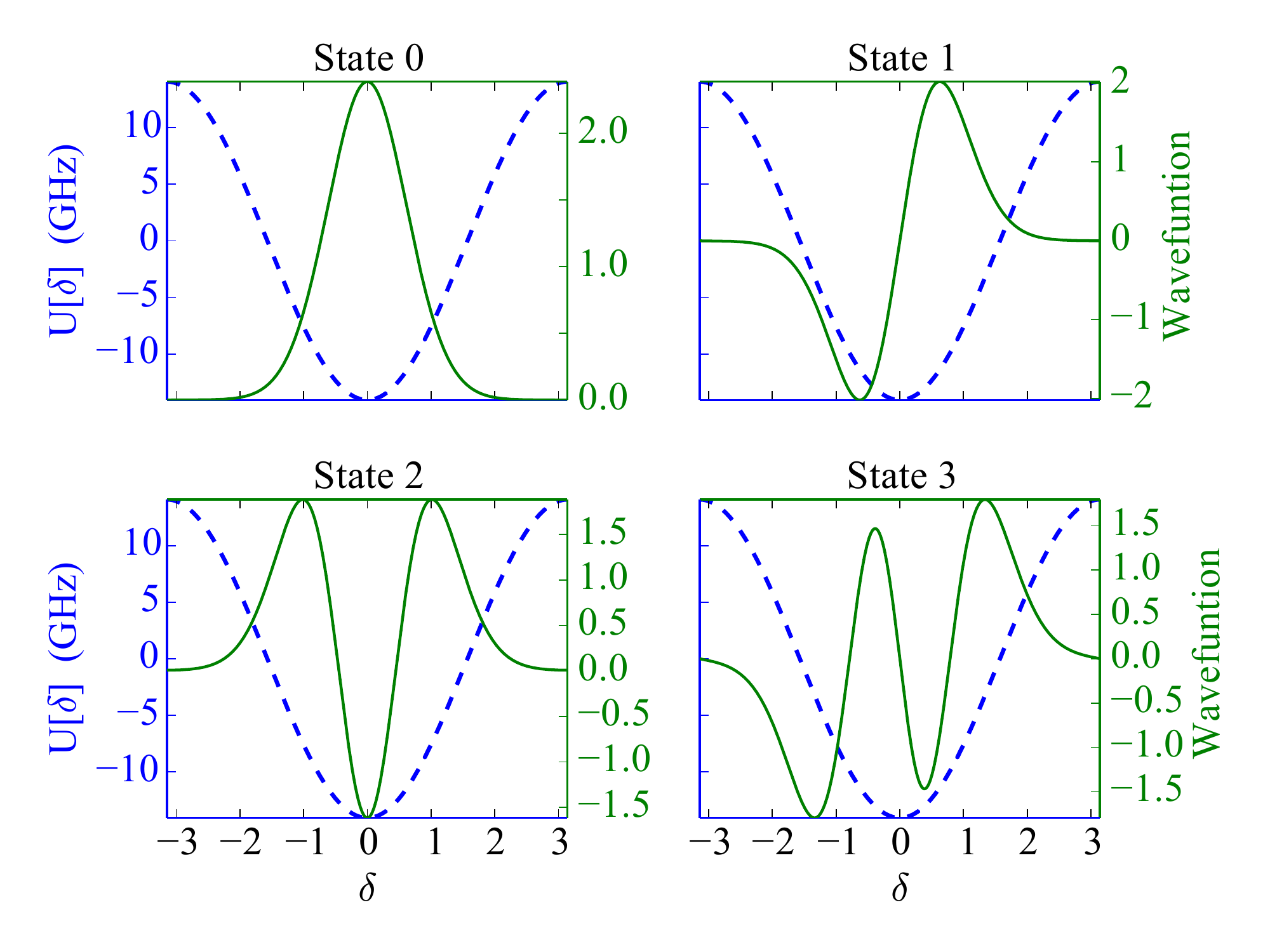}
  \caption{The transmon wavefunctions from simulation in $\delta$-space, overlaid against the Josephson potential $U[\delta]$.  Note that, for this figure, the Hamiltonian was solved at $n_g=0$, which allows for a pure-real energy eigenbasis.}
  \label{fig:wavefunctions}
\end{figure}

The Hamiltonian of the transmon can be written in terms of the gauge-invariant phase $\hat{\delta}$ across the Josephson junction, and its conjugate variable $\hat{n}$, the number of excess Cooper pairs which have tunnelled across the junction, as 
\begin{equation}
    H_T=4E_C\left( \hat{n}-n_g \right)^2-E_J\cos\hat{\delta},
    \label{eq:trans_ham}
\end{equation}
where $E_C$ and $E_J$ are the capacitative ``charging energy'' and the ``Josephson energy'' respectively, and $n_g$ is an effective offset charge \cite{Koch:2007}. This simple, unperturbed Hamiltonian has an analytic solution in terms of the Mathieu functions, though, for later computational purposes, it will be easier to work in the charge basis. The energy spectrum is shown in Fig.\,1(b) and the wavefunctions are displayed in \FigRef{fig:wavefunctions}.

The transmon is coupled to a resonator, which is introduced independently as a harmonic oscillator with frequency $f_c$
\begin{equation}
  H_R=\hbar 2\pi f_c(\hat{a}^\dagger \hat{a}+1/2),
\end{equation}
where $\hat{a} \,\,(\hat{a}^\dagger)$ annihilates (creates) one photon in the resonator.

Finally, a third term includes the dipole coupling
\begin{equation}
  H_D=2e\beta V_\mathrm{rms}\hat{n}(\hat{a}^\dagger+\hat{a})
\end{equation}
between the transmon and the cavity, where $e$ is the electron charge, $V_\mathrm{rms}$ is the RMS voltage of a single photon in the cavity, and $\beta$ is a capacitive divider ratio expressing how much of that voltage the transmon sees.  

Altogether, the quantized circuit is described by the effective Hamiltonian
\begin{align*}
  H&=H_T+H_R+H_D\\
&= 4E_C\left( \hat{n}-n_g \right)^2-E_J\cos\hat{\delta} + \hbar 2\pi f_c (\hat{a}^\dagger \hat{a}+1/2)\\ 
&\qquad\qquad\qquad\qquad\qquad\quad+ 2e\beta V_\mathrm{rms}\hat{n}(\hat{a}^\dagger+\hat{a}).
\end{align*}

The Hamiltonian is rewritten in the basis of the uncoupled transmon states $\ket{i}$ \cite{Koch:2007}, leading to the generalized Hamiltonian which considers higher levels of the transmon
\begin{align}
H = \hbar \sum_{j}{} 2\pi f_j \ket{j}\bra{j} + \hbar 2\pi f_c \hat{a}^\dagger \hat{a} +\sum_{i,j}{} g_{ij}\ket{i}\bra{j}(\hat{a}^\dagger+\hat{a}).
\end{align}
The sum contains the coupling energies:\\
$\hbar g_{ij} = 2e\beta V_\mathrm{rms}\bra{i}\hat{n}\ket{j} = \hbar g_{ij}^*.$

\subsection{Procedure}

The modelling of the energies proceeds as follows.  First, since the interaction involves $\hat{n}$, we numerically diagonalize $H_T$ itself in terms of the $\hat{n}$ basis kets. The relevant Hilbert space of the transmon can then be well represented for our purposes by only the lowest few transmon eigenstates. For the figures and tables in this work, the highest transmon state discussed is state $\ket{8}$.  However, the lowest twenty transmon states were used in simulation, which provides an energy margin of approximately seven times the cavity frequency.

From the resonator Hilbert space, we restrict our attention as well to only the lowest twenty states. Then, representing $H$ on the Kronecker basis of the lowest $H_T$ and $H_R$ eigenstates, we re-diagonalize the entire Hamiltonian and read off the energies.

Of course, this requires knowledge of the various constants in the Hamiltonian: $E_J$, $E_C$, and $g$.  To quickly obtain starting guesses for the energy scales from the frequencies, one can use the approximate relations
\begin{equation}
  h(f_{01}-f_{12})\approx E_C, \quad h f_{01}\approx \sqrt{8E_JE_C}-E_C,
\end{equation}
which can be derived from perturbation theory \cite{Koch:2007}.  Those constants can be refined by fitting to the first two frequencies in the transmon and the dispersive shift of the first level, see Tab.\,I.  Thereafter the model will be able to predict the other frequencies, as well as the other dispersive shifts.
Once the model is fitted, we can easily examine the properties of our system. All fitted and simulated values are given in Tab.\,I.

\subsection{Results}

\begin{figure}[b]
  \centering
  \includegraphics[width=\linewidth]{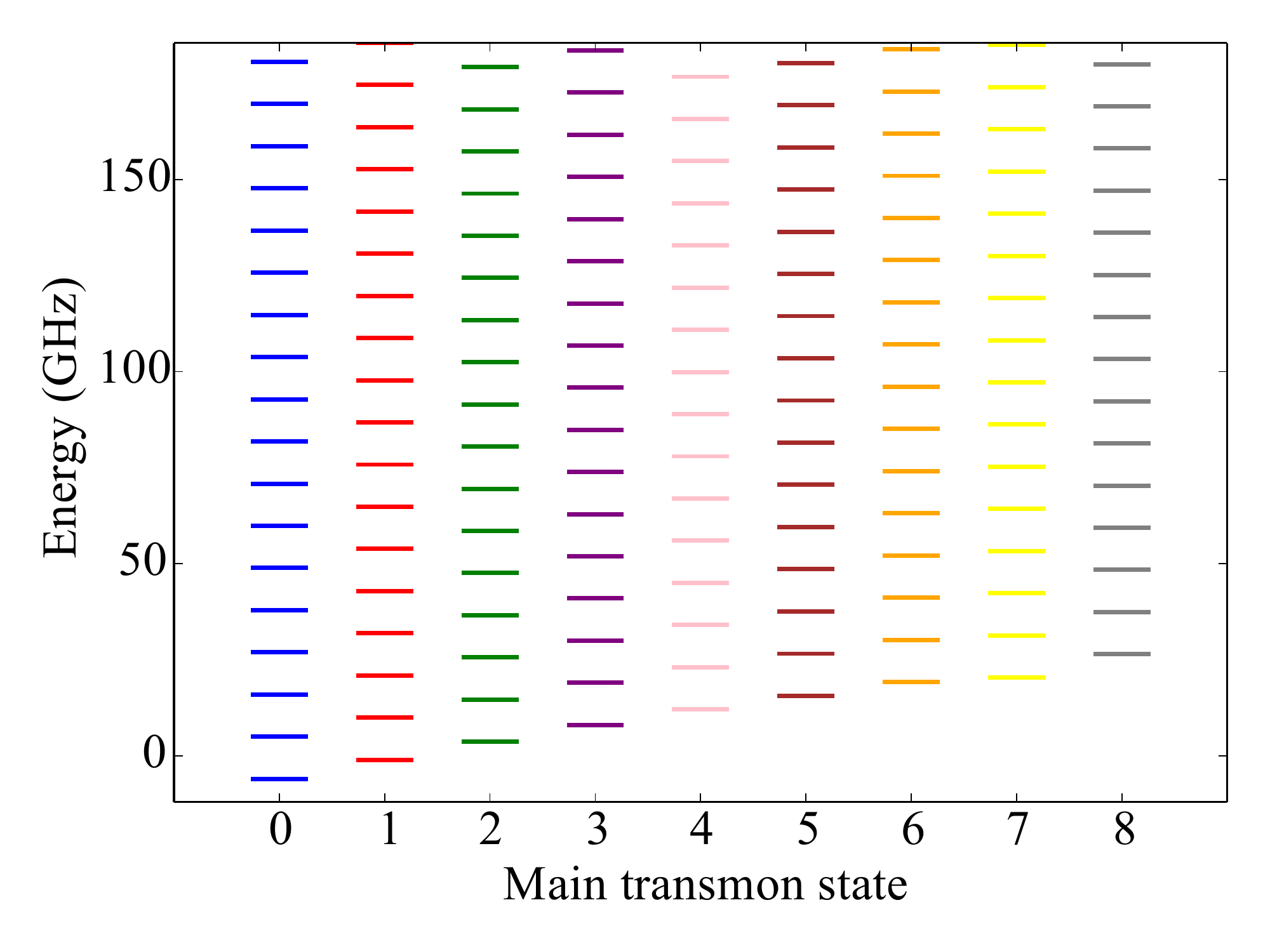}
  \caption{Level scheme for the coupled transmon system up to state $\ket{8}$. We can group the eigenstates of the total system by which transmon state each projects onto the most when the resonator is traced out.  We see that, corresponding to each transmon eigenstate, there is a ladder of total eigenstates, separated by a regular frequency.}
  \label{fig:ladders}
\end{figure}

This simulation provides valuable insight into the aforementioned difficulty of experimentally probing states $\ket{i}$, for $i>2$.  The dispersive readout method assumes that the transmon transitions are in the ``dispersive regime,'' i.e. the detuning of any transmon frequency from the resonator is far greater than the coupling strength, so the resonator only provides a small effective frequency shift to the barely intermixed transmon states \cite{Blais:2004}. The opposite situation, where the resonator frequency is close to a transition frequency, leads to heavy mixing between transmon states, and is known as the ``resonant regime'' \cite{Koch:2007}.

\begin{widetext}

\begin{figure}[t]
  \centering
  \includegraphics[width=16cm]{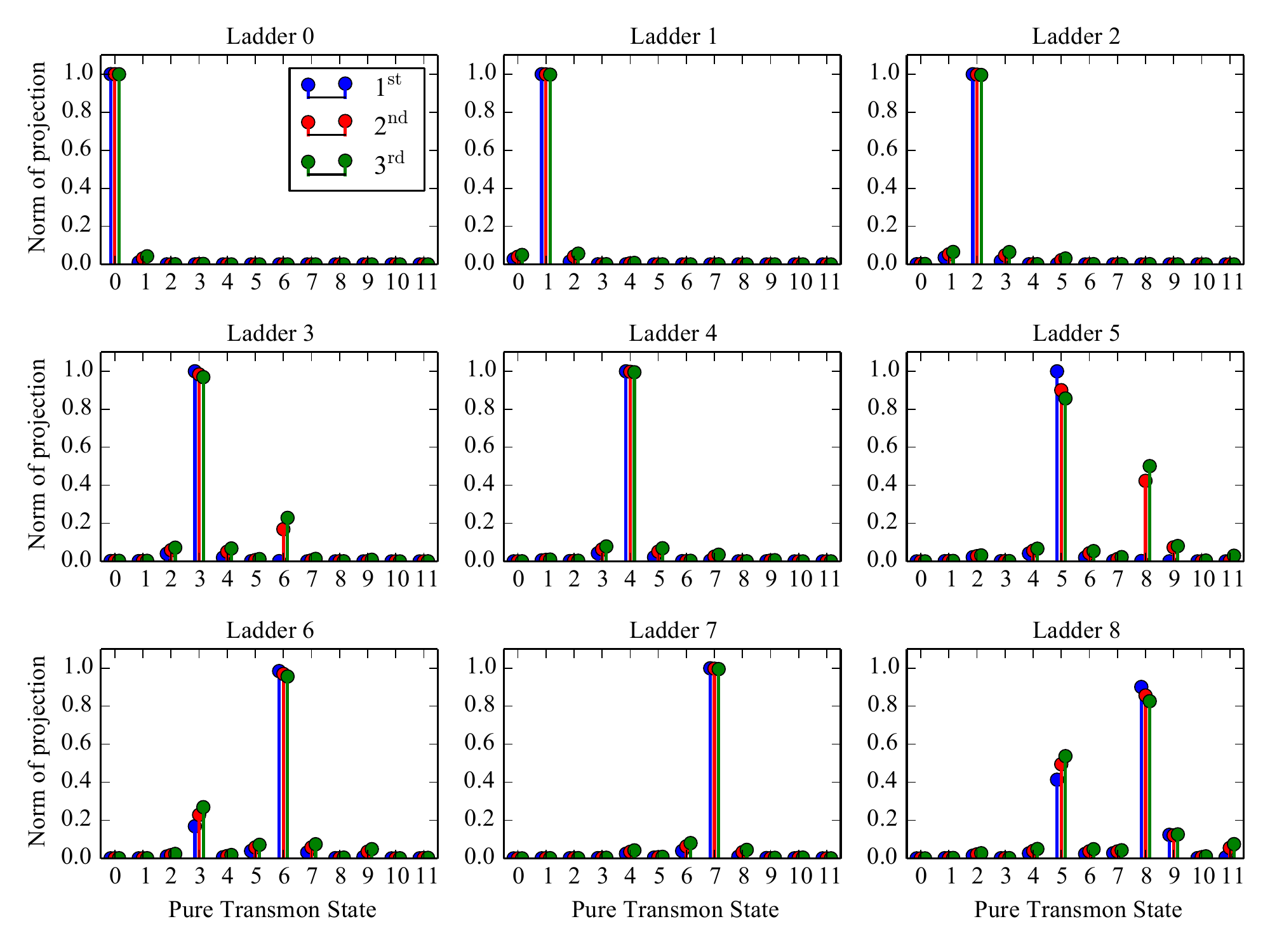}
  \caption{Projections of the first, second, and third states of each ladder onto the pure transmon states with the resonator traced out, per \EqRef{eq:proj}.  We see that the first ladder of states are almost \textit{entirely} transmon state $\ket{0}$.  For instance, if the system were in the lowest eigenstate of Ladder 0 and the experimenter directly measured the transmon state, the experimenter would almost always find 0.  Dissimilarly, states from Ladder 3 have significant projections onto both transmon state $\ket{3}$ and transmon state $\ket{6}$, because the 3-6 transition is resonant with the cavity frequency.  The same can be said of 5-8.}
  \label{fig:projections}
\end{figure}

\end{widetext}

To examine this distinction in practice, we can cluster our diagonalized total eigenstates into different ``ladders'' based on which pure transmon state they most project onto, as in \FigRef{fig:ladders}.  That is to say, ``Ladder 0'' is the set of total energy eigenstates which, when the resonator is traced out, project mostly onto pure transmon state $\ket{0}$.

Explicitly, if $\psi_T^k$ and $\psi_R^l$ represent the transmon and resonator eigenstates respectively, then the projection of a total state $\Psi_\mathrm{TR}$ onto the pure transmon state $k$ is given by
\begin{equation}
 \mathrm{Proj(}\Psi_\mathrm{TR},k\mathrm{)}= \sum_l\bra{\psi_T^k}\braket{\psi^l_R\vert\Psi_\mathrm{TR}} \label{eq:proj}
\end{equation}
and an eigenstate $\Psi_\mathrm{TR}$ is in ``Ladder $i$'' if $\vert$Proj($\Psi_\mathrm{TR},k$)$\vert$ is maximized by $k=i$. The values of $\vert$Proj($\Psi_\mathrm{TR},k$)$\vert$ for the first three states in each ladder are shown in \FigRef{fig:projections}.

If we are in the dispersive regime, we expect to find, for each pure transmon eigenstate $i$, a ladder of total eigenstates separated by some regular frequency $f_c+\chi_i$.  And each total eigenstate in a given ladder should project almost \textit{entirely} onto the same pure transmon state when the resonator is traced out.  Referencing the projections of each total eigenstate, given in \FigRef{fig:projections}, we find this to be the case for the ``well-behaved'' transmon states $\ket{0}, \ket{1}, \ket{2}, \ket{4},$ and $\ket{7}$.  And the spacing of the eigenstates in these ladders is highly regular (only varying on the kHz range in simulation). However, when we examine Ladder 3 in \FigRef{fig:projections}, we see that these states also have a significant projection onto pure transmon state $\ket{6}$, and vice-versa. The same mixing occurs for Ladders 5 and 8. This suggests that these transitions are not well described by the dispersive regime. Indeed, the simulation predicts that the 3-6 transition is a mere 187 MHz detuned from the resonator, and 5-8 is only 91 MHz detuned.\linebreak

A transition $i$-$j$ is in the dispersive regime only if the ratio $|\Delta_{ij}|/g_{ij}\gg 1$, that is, the cavity-transition detuning dominates over the coupling strength. This condition prevents states $i$ and $j$ from mixing.  These ratios are tabulated in \TabRef{tab:rats} for each significant transition, and we see that the ratio is lowest specifically for the noted troublesome transitions.

These low values are of order 1, so these transitions are neither resonant nor dispersive, and our intuition from either regime is sure to break down.  In fact, the eigenstates of Ladder 3 are \textit{not} evenly separated; the spacing varies on the MHz scale (that is, three orders of magnitude more variation than for properly dispersive Ladders in simulation).

Consequently, attempted dispersive measurements on transmon state $\ket{3}$ will not find a stable dispersive shift to associate with the state, but rather a complex, noisy profile representing the interactions of the transmon state $\ket{3}$ and transmon state $\ket{6}$ with the resonator.

However, the experiment was able to circumvent this difficulty and probe state $\ket{3}$ by using a readout scheme with a depopulation sequence which renders it only dependent on the dispersion of states $\ket{0}$ and $\ket{1}$.  In general, future qudit schemes on transmon systems may have to plan around these ``accidental resonances'' which are bound to arise as more transitions come into play.

\begin{table}[h!]
  \centering
  \begin{tabular}{c| c c c c c c c c c}
    \hline
    \hline
    State&0&1&2&3&4&5&6&7&8\\
    \hline
    0&&36.5&&561.5&&&&&\\
    1&36.5&&27.8&&181.1&&&&\\
    2&&27.8&&24.6&&45.5&&&\\
    3&561.5&&24.6&&23.5&&\color{red}5.8&&\\
    4&&181.1&&23.5&&24.1&&40.7&\\
    5&&&45.5&&24.1&&25.3&&\color{red}2.2\\
    6&&&&\color{red}5.8&&25.3&&25.7&\\
    7&&&&&40.7&&25.7&&33.5\\
    8&&&&&&\color{red}2.2&&33.5&\\
    \hline
    \hline
  \end{tabular}
\caption{Compilation of $|\Delta_{ij}|/g_{ij}$ ratios for each transmon transition in the system.  Infrequent transitions for which this ratio is much greater than 1000 (such as parity-forbidden or far off-diagonal jumps) have been omitted for ease of viewing.  We see quantitatively that the 3-6 transition and the 5-8 transition are of order unity, and not safely within the dispersive regime.}
  \label{tab:rats}
\end{table}

\end{document}